\documentclass[preprint,12pt,authoryear]{elsarticle}

\usepackage[utf8]{inputenc}
\usepackage{multicol}
\usepackage{color}
\usepackage{lscape}
\usepackage{xspace}
\usepackage{enumerate}

\title{Chapter 15: Open Questions and Future Directions\\ in Titan Science}

\title{}

\author[label1]{Conor A. Nixon}
\affiliation[label1]{organization={Planetary Systems Laboratory},
             addressline={NASA Goddard Space Flight Center},
             city={Greenbelt},
             postcode={20771},
             state={MD},
             country={USA}}
             
\author[label2,label3]{Nathalie Carrasco}
\affiliation[label2]{organization={Universite Paris-Saclay},
             addressline={UVSQ, CNRS, LATMOS},
             city={Guyancourt},
             postcode={78280},
             country={France}}
\affiliation[label3]{organization={ENS Paris-Saclay},
             addressline={4 Av. des Sciences},
             city={Gif-sur-Yvette},
             postcode={91190},
             country={France}}
             
\author[label4]{Christophe Sotin}
\affiliation[label4]{organization={Laboratoire de Planetologie et Geodynamique, UMR 6112, CNRS, Universite de Nantes},
             addressline={2 rue de la Houssinier},
             city={Nantes},
             postcode={44322},
             country={France}}

\date{Feb 2024}

\begin{document}

\newcommand{\nitrogen}{N$_2$}
\newcommand{\methane}{CH$_4$}
\newcommand{\ammonia}{NH$_3$}
\newcommand{\acet}{C$_2$H$_2$}
\newcommand{\coo}{CO$_2$}
\newcommand{\dg}{$^{\circ}$}

\maketitle


\section{Introduction: Current Sources of Knowledge}

In the preceding 14 chapters, a detailed description of the current state of knowledge about Titan in the era of Cassini-Huygens has been laid out. In this chapter, we finally take a step back from the details and look at the big picture: how did we arrive at this current state of knowledge? What are the large topical areas that remain mysterious? And how can we attempt to answer these questions via future investigations?

First let us recap at the high level how our constraints on Titan have been collected. There are primarily four techniques through which knowledge has been gained:

\vspace{4mm}

\emph{Telescopic observations:} Titan was first discovered by means of telescope by Christiaan Huygens in 1655, and until the post-WW2 space age, this remained the only direct means of gathering data about Titan. Seminal telescopic findings about Titan include the discovery of the first atmospheric gas, methane, by Gerard Kuiper in 1944 \citep{kuiper44}; the discovery of carbon monoxide \citep{lutz83} which helped to resolve questions about the atmospheric bulk composition and density; the active radar measurements of surface albedo that showed a lack of global ocean \citep{muhleman90}; the first imaging of the surface \citep{smith96}; the discovery of clouds \citep{griffith98}. Further descriptions of telescopic observations of Titan throughout the pre-Cassini and Cassini era can be found elsewhere in this volume: see especially Chapter 7 for details on observations of the atmosphere and Chapter 11 for surface observations.

\vspace{2mm}

\emph{Laboratory Work:} Laboratory research has filled many gaps in knowledge left by remote sensing and the limited in-situ investigations performed to date. This especially includes the investigations of analog haze materials (`tholins') \citep[see review by][]{cable12} which has provided significant insights into the atmospheric chemistry, the nature of haze materials, and the potential for alteration of these grains especially in hydrolysis reactions that can lead to amino acids \citep{khare86, neish09}. Moreover, laboratory gas and ice spectroscopy provided important insights to decipher the observations made during the Cassini space mission \citep{sung13, sung18, sung20, bernath21, bernath23, anderson18}. Reaction rates and pathway measurements (kinetics) have informed our knowledge of chemical processes taking place in Titan's atmosphere \citep{balucani10,kaiser15,dutuit13,gans13,carrasco22}.  Clathrate and co-crystal formation investigations \citep{cable21a}, and aeolian, pluvial and fluvial erosion experiments  \citep[e.g.][]{litwin12,burr15a, burr15b,yu17a, yu17b,yu18, yu20, burr20, maue22, mackenzie23, hirai23} deeply improved our understanding of the observed features at Titan's surface. 

\vspace{2mm}

\emph{Modeling:} Some problems must be tackled computationally, notably dynamical models of solar system evolution and accretion scenarios that provide insights into Titan's origins in the context of the Saturn system \citep[][and Chapter 4]{mousis02a, mousis02b}. However, models specializing in particular areas of Titan science have proved immensely valuable at explaining observations and posing further questions. These include: upper atmosphere sputtering and escape models \citep[][and Chapter 6]{lammer93, lammer98, mandt12, johnson09, gu19, shematovich03, snowden21, johnson94, krasnopolsky16, tseng08};  
bulk atmosphere climate and dynamical models including GCMs (Global Circulation Models) \citep[][and Chapter 8]{delgenio93, hourdin95, mitchell08, friedson09, lebonnois12, tokano13, lora15, lora19, chatain22}; 
mesoscale and regional models \citep[][and Chapter 8]{tokano09, charnay15, mitchell11, rafkin20, spiga20, chatain22}; 
photochemical models \citep[][and Chapter 7]{yung84, wilson03, lavvas08a, lavvas08b, krasnopolsky09, krasnopolsky10, vuitton19, loison15, dobrijevic14}; 
microphysical models of cloud droplet and haze formation \citep[][and Chapters 7 and 8]{barth04, barth06, burgalat14, larson15}; 
surface models including erosion, dune formation, river branching and more \citep[][and Chapters 9--11]{tokano08, barnes15, charnay15, burr13a, birch23}; 
crustal tectonic models \citep[][and Chapter 13]{mitri08a, liu16b, burkhard22, schurmeier23};  
impact models \citep[][and Chapter 3]{griffith95, artemieva03, kress04, korycansky11, zahnle14, crosta21}; 
interior structure models \citep{grasset00, tobie06, mitri08b, tobie09, sohl14, journaux20} and oceanographic models \citep[][and Chapter 12]{tyler08, tyler14, soderlund19, vincent22, soderlund23}.

\vspace{2mm}

\emph{Spacecraft Missions:} The missions that have reached the Saturn system include Pioneer 11 (1979), Voyager 1 (1980) and Voyager 2 (1981); and Cassini-Huygens (2004-2017) \citep{kohlhase77, matson02, lebreton02}. These missions have built on each other and each of the generations of spacecraft has successively provided much greater detail than the previous. Of these, only Voyager 1 and Cassini-Huygens have been actively targeted to visit Titan, and while Voyager 1 provided an exciting first look at Titan, sufficient to stimulate excitement for a follow-on mission, it was Cassini-Huygens that first revealed Titan in detail. The results from Cassini-Huygens have been extensively described in this book (see Chapter 2 for mission overview).

\vspace{2mm}

At the end of this chapter we will return to explore how these four techniques may be used to answer the large, high-level open questions in Titan science. First, we will describe the principal open questions themselves.

\section{Open Questions}

We have attempted to distill the very large number of possible future inquiries of Titan into a relatively concise list of twenty high level questions,\footnote{Inspired by David Hilbert's 23 unsolved mathematical problems of 1900 \citep[republished in][]{hilbert00}.} each of which of would necessarily entail a multitude of more specific investigations and studies. While this list may not encompass all possible open questions, and is divided into topics according to our preference and not in any way uniquely, we believe that it does however span a wide range of the most intriguing topics about Titan, and may form some sort of guide especially for those embarking into Titan studies for the first time. See Fig.~1 and Table~1.

\begin{figure}[htbp]
\centering
\rotatebox{90}{\includegraphics[width=1.3\textwidth]{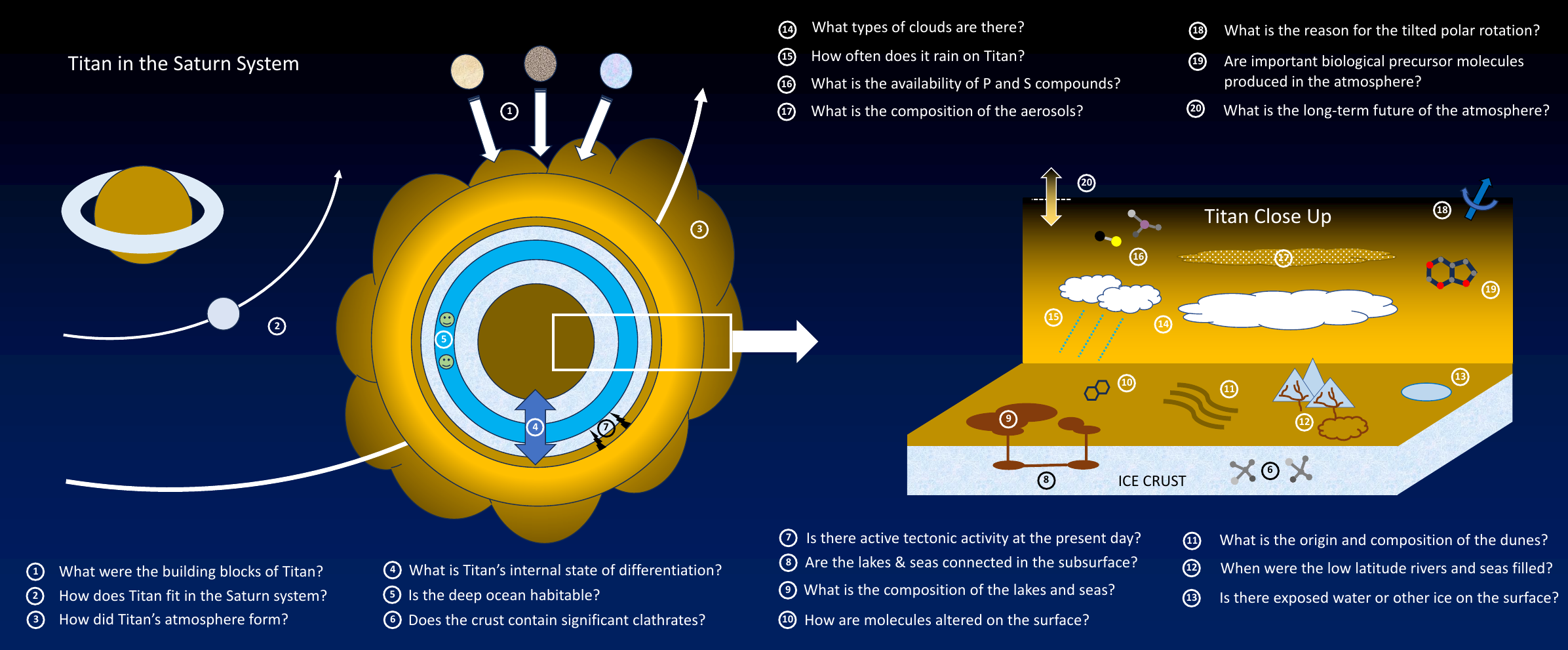}} \\[0.2cm]
\caption{Open questions in Titan science.}
\label{fig:questions}
\end{figure}

\begin{landscape}
\begin{table}
\begin{center}
\caption{Key Open Questions in Titan Science}
\includegraphics[width=1.5\textwidth]{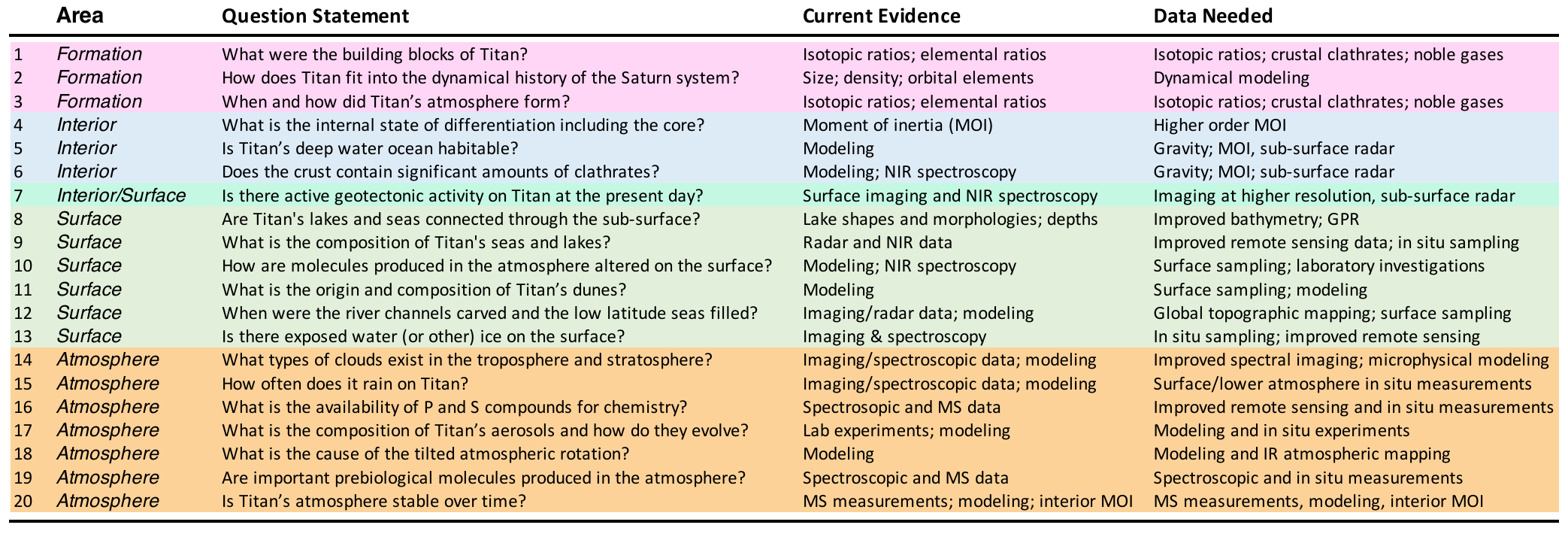} \\[0.2cm]
\end{center}
\begin{small}
Abbreviations: 
MOI = moment of inertia. MS = mass spectroscopy. 
NIR = near infrared. GPR = Ground penetrating radar.
\end{small} 
\end{table}
\end{landscape}

\subsection{Questions about Titan's formation (1-3)}

\vspace{5mm}
\emph{1) What were the building blocks of Titan?}

Our current sparse constraints regarding Titan's interior permit multiple interpretations of its core composition. The moment of inertia (C/MR$^2 \sim$0.34) from gravity measurements by Cassini \citep{iess10, iess12, durante19} can be modeled in several ways, including a bimodal model with a mixture of low density and high-density components \citep{neri20}, or a more uniform composition lacking in significant amounts of iron \citep{vance18}. 

We must also explain the presence and bulk composition of the atmosphere, mostly made of nitrogen and methane. The current favored paradigm for the origin of nitrogen is photolysis of ammonia \citep{mandt14}, while the presence of methane is still a subject of intense debate (see Q3), with multiple possibilities including trapping as primordial clathrate \citep{mousis15, tobie06} and generation in the interior \citep{glein15, miller19}.  

Other atmospheric quantities can in principle provide constraint on building blocks, such as the ratios of stable isotopes (D/H, $^{12}$C/$^{13}$C, $^{14}$N/$^{15}$N, $^{16}$O/$^{18}$O, etc) \citep{pinto86,wong02,jennings08,mandt09, nixon12b, mandt12}, stable noble gas abundances ($^{20}$Ne and $^{22}$Ne, $^{36}$Ar and $^{38}$Ar) and the abundances of radiolytic elements (e.g. $^{40}$Ar) \citep{glein15, miller19}, as described in Chapter 3. If unaltered from formation, or with adequate knowledge of alteration processes, the present-day ratios and abundances constrain the values at time of formation through comparison to ratios in other primitive materials such as comets and meteorites \citep{mandt14, glein15, miller19}.

However, making inferences about building blocks from present-day isotopic ratios in the atmosphere is problematic, since it requires assumptions about how the atmosphere has evolved \citep{owen09, nixon12b, mandt12, wong15}. Similarly, making interpretations from noble gas abundances is reliant on assumptions about the amount of internal differentiation and outgassing.

More robust isotopic ratios could be derived from measurements of crustal material, with the D/H in water - likely the main component of the crust - being particularly informative \citep{coustenis09}. Measurements from Dragonfly DraMS may be helpful here \citep{barnes21, grubisic21}. In addition, better measurements of the gravity field will yield more precise values for the higher degrees of MoI ($J_2$...) and the Love numbers ($h_2$, $k_2$...) which can be used to make inferences about the distribution of mass and the rigidity of the ice shell \citep{iess10, iess12}. 

With such future measurements, we will be better positioned to find answers to our questions about the primordial building blocks of Titan.

\vspace{5mm}
\emph{2) How does Titan fit into the dynamical history of the Saturn system?}

The Saturn system appears less regular than the Jovian system with its four large satellites and multiple smaller, apparently captured bodies. In contrast, Saturn's moon system is dominated by the giant Titan, with the other mid-sized moons much less massive in comparison, adding in total to less than 5\% of Titan's mass. 

Recent numerical work has pointed to a potentially young age for the Saturnian moons interior to Titan (i.e. closer to Saturn) \citep{cuk16, lainey17, wisdom22, teodoro23}, although this has been disputed based on surface age estimates \citep{wong23}. In addition, the dynamical ages depend on the initial value of Saturn's tidal quality factor, which is a poorly known parameter. \citet{neveu19} argue for a large initial value for this parameter in order to explain the observations of each moon, which leads to a formation age similar to that of Saturn, with the exception of Mimas, which may have formed later.

Also, Saturn possesses the only icy ring system, which appears unstable over time \citep{kempf23} and may have been formed relatively late by disruption of an inner moon \citep{dubinski19}. These strands of evidence point to possible evolution of the Saturn system at recent times \citep{crida19}, and the role of Titan in this drama remains unclear (see Chapter 4). 

Factual evidence to constrain models comes from the masses and orbital parameters of the satellites, the mass and composition of the rings, and the age of the moons as implied from surface crater counts. Improvements to our models can come from more sophisticated modeling, comparison to other satellite systems (including exoplanet moon systems) and especially better understanding of cratering histories and rates in the outer solar system, which remain significantly uncertain at the present time \citep{neish12, nesvorny23}.

\vspace{5mm}
\emph{3) When and how did Titan's atmosphere form?}

The Voyager 1 flyby of Titan in 1980 provided the first accurate measurements of the atmospheric composition, confirming suspicions that N$_2$ was the dominant gas, and finding \methane\ to be the next most abundant constituent. Subsequent photochemical modeling \citep{strobel82, yung84} revealed a surprising corollary: the current depletion rate of methane means that the present atmospheric inventory of \methane\ will be entirely lost in a timescale of a few 10's of Myr: much less than the age of the solar system. The age may be prolonged if methane is being actively resupplied to the atmosphere, such as by clathrate displacement \citep[Chapter 3][]{choukroun12}.

Therefore, either Titan's methane is young and recent, or else we are observing the extinction of its last remnants at the end of a long history. However, the lack of sufficient hydrocarbon products on the surface makes the latter hypothesis difficult to sustain \citep{lorenz08a, wilson09}. As an alternative, it has been suggested that the atmospheric methane may have originated in a relatively recent event of crustal destabilization \citep{tobie06}.

Attempts have been made to estimate the age of methane's presence in the atmosphere through isotopic evolution by Raleigh distillation \citep[Chapter 3][]{mandt12, nixon12b}. However, the significant error bars on present-day methane isotopic ratios, combined with uncertainties about the original D/H and $^{12}$C/$^{13}$C ratios make it difficult to place quantitative values on the methane age. Another possible means of calculating the atmospheric age is to add up the surface hydrocarbon deposits and compute how long these would take to accumulate from atmospheric photochemistry, yielding upper bounds of 630 Myr \citep{sotin12} and 730 Myr \citep{Rodriguez11}.

More precise values for the present day $^{12}$C/$^{13}$C, D/H and other isotopic ratios are required, and in addition some improved knowledge of their initial values, perhaps through crustal measurements, could be used to better infer the atmospheric age. Another route is a better knowledge of the amounts of hydrocarbons resulting, including both on the surface and any sequestered in the subsurface. Substantial progress on the surface hydrocarbon inventory has been made \citep{lorenz08a, sotin12, rodriguez14}, but knowledge of the sub-surface inventory remains speculative. Future ice-penetrating radar and/or seismic measurements may help to determine the porosity and density of the crust \citep{mitri21, rodriguez22}, providing better constraint on its composition.

\subsection{Questions about Titan's interior (4-6)}

\vspace{5mm}
\emph{4) What is the internal state of differentiation including the core?}

Strong evidence for differentiation within Titan comes from gravity measurements, which provide values for the Moment of Inertia (MoI) and the degree 2 tidal Love number. The MoI suggests that Titan is differentiated into a hydrosphere and a refractory core. However, Titan does not appear to have a iron-rich core like Ganymede, whose MoI value suggests a higher degree of differentiation. 

Analyses of gravity data  during and shortly after the mission resulted in estimated values for Titan's tidal Love number $k_2$ of $\sim$0.6 \citep{iess10, iess12, durante19}, although with a large error range. Such a high value of $k_2$ implied not only the presence of an ocean that decouples the icy crust from the solid interior \citep{mitri14b}, but moreover that the ocean is much denser than pure water. 

A more recent reanalysis of the gravity data \citep{goossens2024low}, including a more elaborate treatment of tidal effects, has challenged these earlier findings with a much lower estimate for $k_2$ of 0.375$\pm$.0.06. This lower value implies a much lower ocean density of 1091$\pm$107~kg~m$^{-3}$, compared to the previous estimate of 1300~kg~m$^{-3}$.
A major question, which also affects the habitability of the ocean, concerns the presence and properties of a high pressure water ice layer (or layers) at the base of the ocean \citep{journaux17, journaux20, kalousova20a}, affecting the exchange of core materials with the ocean \citep{miller19}. See also Q5.

Future measurements that can better elucidate the interior layering include further gravity data, measurements of the surface deformation during the orbit around Saturn, and seismic observations. Note that we cannot try to measure an induced magnetic field at Titan to learn about its ocean, as we can for the Galilean moons of Jupiter. This is because Saturn's magnetic field is perfectly aligned with its spin axis, which prevents the formation of an induced field that would be created by the presence of a conductive fluid (salty ocean) in a time-varying magnetic field.

\vspace{5mm}
\emph{5) Is Titan's deep water ocean sterile?}

That Titan is an ocean world - in the sense of having a deep interior water ocean - is evidenced by measurements of the degree 2 tidal Love number $k_2$ \citep{iess12, durante19, goossens2024low}. However, what is much less certain is the geochemical state of the ocean and whether it contains sufficient organics and minerals to permit life to exist \citep[see review by][and also Chapter 14 of this book]{journaux20}.

Several recent papers have examined the possible interaction of the ocean with the surface and the core. \citet{kalousova20b} show that the presence of clathrates (Q6) will thin the crustal ice shell, allowing greater chance of convection. The analysis of \citet{fortes07} suggests that ammonium sulfate released from the core will likewise increase the chances of crustal ice convection. 

Work by \citet{miller19} suggests that the core of Titan is warm and rich in refractory organics, which decompose to release gases that percolate through the high pressure (HP) ice layer to reach the ocean and eventually to the atmosphere. \citet{journaux17} considered the effect of salt dissolution on the mobility of the high-pressure ice layer, showing that convection is possible but not certain. More recently \citet{kalousova20a} further modeled convection within the HP ice layer, concluding that convection is likely. 

It is important that one or both of these interactions (core-ocean, and surface-ocean) is viable, to permit significant nutrients to enter the ocean to allow the sustenance of life \citep{fortes00}. A pessimistic estimate of the surface contribution of organics to the ocean was recently obtained by \citet{neish24}, although many uncertainties remain.

Improved constraint on the ocean-core and ocean-crustal boundaries may be achieved by more accurate measurements of noble gases in the atmosphere, as well as interior measurements (gravity field) and modeling of the thermo-chemical evolution of Titan's refractory core, including use of improved data on high-pressure water ice phases.

\vspace{5mm}
\emph{6) Does the crust contain significant amounts of clathrates?}

Clathrates - minerals wherein a crystalline cage of one substance traps an atom or molecule of a different substance - are well known on Earth and are suspected to play an important role in the formation and volatile history of solar system bodies \citep{mousis15}. Methane and ethane clathrates are stable at Titan's surface conditions \citep{lewis71}. Their presence and prevalence in Titan's crust, especially methane clathrate hydrates, has been a key puzzle piece required for our understanding of Titan's interior evolution \citep{lunine87, tobie06}. Clathrates have been proposed to play a dizzying array of roles, including:

\begin{itemize}
\item {\bf Methane outgassing:} Several models have proposed that methane - originally trapped in clathrates - may be later released to replenish the atmospheric methane inventory \citep{lunine87}, ameliorating the problem of \methane\ depletion in short geological timescales. Any injection of ammonia will accelerate the destabilization process \citep{choukroun10}, or alternatively methane may be displaced by ethane produced in the atmosphere by photochemistry \citep{choukroun12}.
\item {\bf Trapping of noble gases:} Studies suggest that noble gases such as Kr and Xe may be efficiently sequestered in surface clathrates, explaining the lack of detection of these gases in the atmosphere \citep{thomas07, thomas08, mousis11}. However, photoionized xenon is also suggested to be more efficiently trapped into the organic haze at high altitudes than other noble gases \citep{hebrard14}.
\item {\bf Crustal thinning:} The insulating effect of a methane clathrate layer at the top of the icy crust has been proposed to lead to a thinner ice shell that is more likely to allow surface-ocean interaction and exchange \citep{kalousova20b}
\item {\bf Sequestration of lake materials:} The relatively rapid formation of clathrates, even under cryogenic conditions, implies that significant sequestration of liquids such as ethane likely occurs in the beds of Titan's seas and lakes \citep{vu20b}.
\end{itemize}

The presence or absence of clathrates has a significant impact on the structure of Titan's interior layering \citep[][and Chapters 10--14]{lunine87, choukroun10, mousis14, davies16, marusiak22}, with implications for its overall history.

For these reasons (and others), determining the prevalence, distribution and types of clathrate present in the crust is a subject of intense importance for understanding Titan. Seismic measurements by Dragonfly may provide a means of direct measurement, since methane clathrate in the crust will attentuate surface waves to a greater degree than pure water ice \citep{marusiak22}. Radar backscatter measurements are also sensitive to the dielectric properties of the surface material, and can be effectively used to constrain composition \citep{thompson90, lorenz98, lorenz03, wye07, paillou08, legall16}. 

\vspace{5mm}
\emph{7) Is there active geotectonic activity on Titan at the present day?}

Ongoing tectonism in Titan's crust, with surface manifestations such as `cryovolcanoes', was postulated in the era following the Voyager 1 encounter \citep{kargel91, kargel92, lorenz95}, primarily based on the need to explain the long-term presence of methane in the atmosphere (see Q3, Q20). However, the higher density of liquid water relative to water ice makes cryovolcanic outflows very unlikely unless overpressure is present, for example due to the global cooling of the icy moon in the absence of a high-pressure ice layer.  

A surface feature seen by VIMS early in the Cassini mission was initially hypothesized to be a signature of lobate flows \citep{sotin05}. However, later observations with RADAR showed that the morphology of the feature was quite bland and indistinguishable from the surroundings at centimeter wavelengths \citep{hayes08}. 

\begin{figure}[htbp]
\centering
\includegraphics[width=1.0\textwidth]{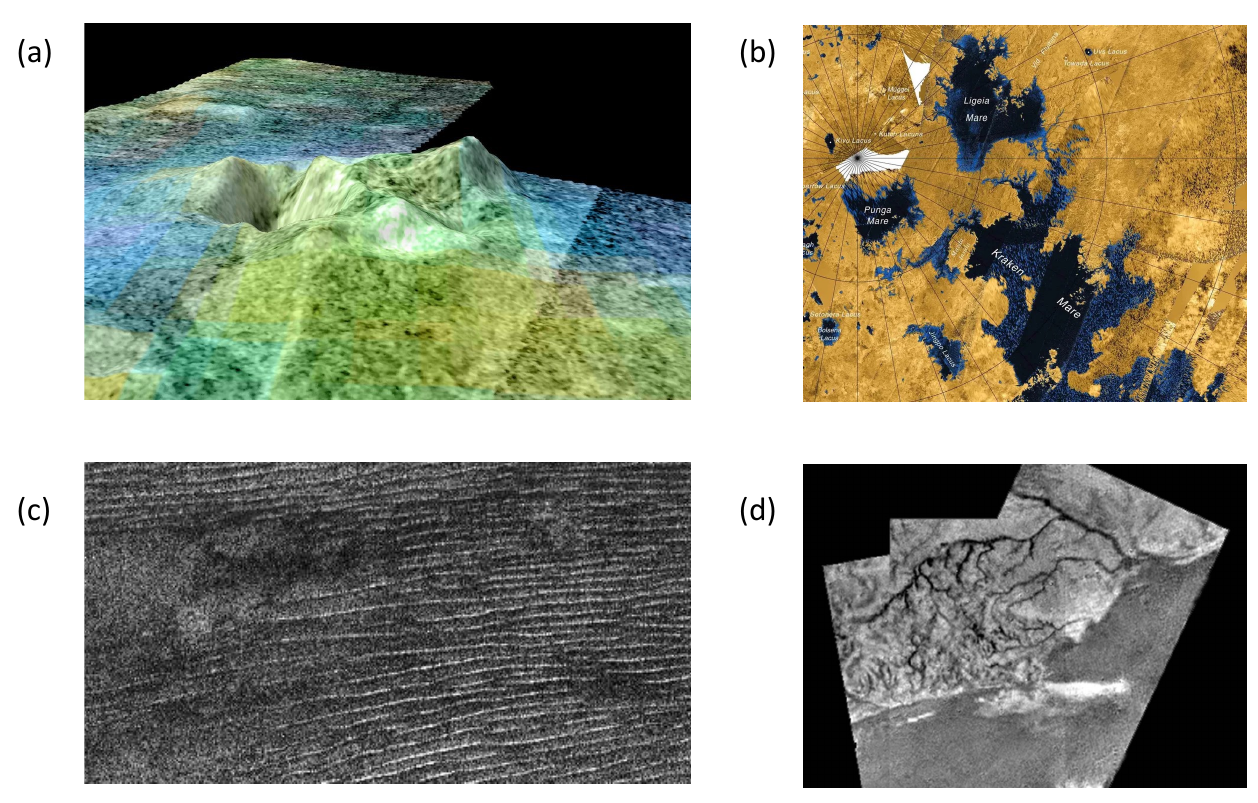} \\[0.2cm]
\caption{Views of Titan's surface. (a) Cassini topographic reconstruction of Doom Mons from Cassini radar SARtopo data - NASA image PIA13695; (b) Cassini colorized radar map of Titan's northern hemisphere seas - NASA image PIA17655; (c) Cassini radar image of Titan dunes in the Belet `sand sea' - NASA image PIA15225; (d) Huygens DISR descent image of low-latitude fluvial networks - NASA image PIA07236. Image credits: NASA/JPL-Caltech/ESA/ASI/USGS/University of Arizona.}
\label{fig:surface}
\end{figure}

The best candidate cryovolcanic region at the end of the mission is Sotra Patera,  which consists of a prominent depression (possible cryovolcanic caldera) called Sotra Patera, a high mountain Doom Mons (Fig.~\ref{fig:surface}(a)), and a lobate flow,  Mohini Fluctus \citep{lopes13, lopes17, lopes19, solomonidou16, wood20}. However, the low-resolution of the RADAR topography makes its exact nature uncertain, and no geophysical model has been proposed to explain how it may have arisen. A review of the evidence for cryovolcanic surface features is given in \citet{lopes13}.


Cryovolcanoes aside, evidence for other types of tectonism is stronger. Cassini SAR imagery has many topographical prominences interpreted as mountain chains, contractional ridges, and similar \citep{barnes07, radebaugh07, mitri10, lopes10, jaumann10, cook15, liu16a, lalich22}. At the southwestern boundary of the mysterious Xanadu region,  evidence for several types of faulting (normal, thrust and strike-slip faults) has been proposed \citep{matteoni20}. Reviews of the evidence for various types of tectonism can be found in the literature \citep[e.g.][]{lopes07, jaumann10, lopes19}, as well as in Chapters 9 and 13 of this volume.

However, the fact that these features mostly appear heavily eroded and there is no evidence of present-day changes makes the timing of their origin uncertain and leaves open the question of whether present day crustal activity is occurring  \citep{solomonidou13}. These ambiguities have led some authors to pronounce Titan geophysically dead \citep{moore11}. However, in view of the low spatial resolution and incomplete nature of the present surface mapping of Titan, reaching a definitive conclusion seems premature.

The principal impediment to gathering the needed high-quality imagery and spectroscopy of the surface, and searching for temporal changes in surface morphology is the presence of the thick atmosphere, which causes several difficulties. First, its haze absorbs and scatters at visible and many near-IR wavelengths, making it difficult to image at shorter wavelengths. In the mid-infrared, emission is dominated by the warm atmosphere, while at longer RADAR wavelengths the problem becomes simply the decreased spatial resolution. Moreover, the atmosphere prevents close flybys at altitudes less than $\sim$1000~km which would allow closer examination.

While Dragonfly will provide a more detailed picture of a small region of Titan's surface, mostly covered in dunes, it will not provide a high quality global map that appears to be required as the next step to unraveling the global geophysics of Titan's crust. A mapping polar orbiter equipped with multi-spectral passive and active surface imaging capability, as well as the ability to conduct sub-surface sounding (e.g. ice-penetrating radar, IPR) would allow us to much more clearly discern whether Titan is an active world at the present day, or not. A long-duration balloon mission could provide regional coverage at high resolution \citep{coustenis09, tobie14}. 

\subsection{Questions about Titan's surface (8-13)}

\vspace{5mm}
\emph{8) Are Titan's lakes and seas connected through the sub-surface?}

The general hydrology of Titan has been described previously in this book (see Chapter~10), so we focus here on the particular question of whether there is subsurface connection between the exposed liquid tracts.

Altimetry measurements by Cassini's RADAR indicate that the surfaces of the major seas (Kraken, Ligeia and Punga) fall on a uniform equipotential surface - interpreted as evidence that they could be connected at a subsurface level \citep{hayes17}. The same work found that apparently dried up lakebeds within the same drainage basins are typically found at higher elevations, adding further weight of evidence for a consistent methane `water table'. However, it appears that Ligeia and Kraken are already connected by a network of surface channels that could help to homogenize their levels \citep{lorenz14b}, and the other apparent similarities in sea surface levels may be coincidental.

Understanding the distribution of liquids on Titan's surface and in the subsurface has important ramifications not just for regional lacustrine studies, but for understanding of the overall climate of Titan, and its seasonal and secular trends \citep{tokano19}. Hydraulic connectivity of Titan's lakes and seas affects our understanding of their distribution, and the creation of surface morphological features through rainfall such as playas, arroyos, labyrinths and karsts \citep{burr13b, perron06, malaska20, solomonidou20}. 

On the global scale, the ability of the atmospheric methane to be buffered by and replenished from the subsurface is one of the most important open topics influencing our understanding of the long-term stability of the atmosphere (Q20). Methane may be released from clathrate in the crust and seep into the methane `water table' and eventually to the surface, providing a resupply to the atmosphere to counter methane loss \citep{choukroun10, choukroun12, mousis14, davies16}.

The subsurface hydraulic connectivity can be investigated in future by improved radar or lidar altimetry over the lakes and seas \citep{lorenz08b}, and especially monitored over seasonal timescales to search for any predicted changes \citep{mitri07, tokano09, tokano19}. Measurements of the subsurface by long-wavelength radar and seismic probing may also be productive. Finally, a direct measurement of the sea surface height and seabed topography can be made by a floating probe or submarine \citep{stofan13, oleson15}

\vspace{5mm}
\emph{9) What is the composition of Titan seas and lakes?}


The distribution of contemporary polar seas (Fig.~2b) is not uniform, with only one significant liquid body being seen in the southern hemisphere at the present day (Ontario Lacus) \citep[][and Chapter~10]{turtle09}. In addition, Ontario's composition appears to be substantially different from the large northern seas, being apparently composed mostly of ethane \citep{brown08}, significantly less volatile than the methane-rich northern seas \citep{mastrogiuseppe19}. \citet{lorenz14b} has further suggested than the increased rainfall at more northerly Ligeia leads to a more methane rich composition ($\geq80$\%) than Kraken ($\sim$60\%).


Theoretical work by Cordier and others \citep{cordier09, cordier12, tan13, malaska14, cordier21} has sought to quantify the expected composition of the lakes including both major and minor constituents, based on predictions of condensation rates of organics from the atmosphere.

The lake/sea composition is important from a habitability standpoint, since it has been hypothesized that prebiotically relevant organic structures can form in the lakes \citep{stevenson15, rahm16} from molecules such as HCN and acrylonitrile, which are known to exist on Titan \citep{palmer17, thelen19b}.

Our knowledge of the bulk composition of the seas may be refined in future by improved spectroscopy in the near-IR and radar loss tanget measurements at multiple wavelengths, along with continuing developments in lab experiments and theory. However, to truly determine the fractions of minor constituents dissolved in the lakes will ultimately require direct measurements through a probe sent to sample the lakes \citep[e.g.][]{stofan13, rodriguez22}.

\vspace{5mm}
\emph{10) How are molecules produced in Titan's atmosphere altered on the surface?}

Titan's atmosphere produces a wide variety of organic molecules \citep{yung84, waite05, waite07, vuitton19} which condense in the lower stratosphere and accumulate on the surface. Simulations in the laboratory have shown that these complex organics include many molecules known to be fundamental ingredients for biology on Earth \citep[see review by][]{cable12}.

An important limitation to the variety of chemicals produced in the atmosphere was thought to be the lack of oxygen involvement in the photochemistry, however lab experiments have shown that carbon monoxide can take part in reactions \citep{horst12}, leading to amino acids.

Moreover, molecules produced in the atmosphere may also be altered once on the surface, e.g. by hydrolysis during episodic melting induced by impacts \citep{neish08, neish09, neish10, poch12, cleaves14, crosta21} - see also Chapter 13. Laboratory simulations have shown that hydrolysis of pure H$_x$C$_y$N$_z$ substances also leads to amino acids \citep{khare86}.

Some evidence for surface alteration exists from the detection of benzene on Titan's surface through near-IR spectroscopy with Cassini VIMS \citep{clark10}. This was a surprising result since the same study found no evidence of hydrocarbon ices that are expected to be more abundant, such as acetylene. A possible explanation was offered that acetylene may be modified (cyclized) into benzene under certain conditions, especially in the presence of catalysts, for example during an impact.

Besides small molecules of astrobiological relevance, it is possible that the vast equatorial dunefields of Titan consist of atmospheric residues, perhaps reprocessed on the surface as suggested by \citet{barnes08} (see next section).

The Dragonfly mission, targeted to land on Titan's low-latitude dunes, is ideally suited to investigate the surface composition, and the relation between atmospheric condensates and surface molecules and particulates. Additional laboratory work will be important to verify and interpret the Dragonfly data.

\vspace{5mm}
\emph{11) What is the origin and composition of Titan's dunes?}

Titan's equatorial dune fields (or `sand seas', Fig~2c) were discovered during the earliest parts of the Cassini mission by RADAR imaging \citep{lorenz06} and subsequently investigated in detail. These are known to cover $\sim 17$\% of the planet, mostly at latitudes between $\pm$30\dg , with a combined mass of $\sim$230,000 GT \citep{rodriguez14, lopes20}.

Multiple hypotheses exist for the formation of dune material \citep{barnes15}, including the `top down' processes of sintering and flocculation, and the `bottom up' processes of lithification and evaporite formation. However which process or processes dominate the particle formation remains as yet unknown. Once created, the particles form the large-scale, mostly longitudinal dune structures due to prevailing surface winds \citep{tokano08}.

Key questions related to the dunes include: are they formed through `top down' or `bottom up processes'? How long did they take to reach their present configuration, and is this stable over time? \citep{rubin09, savage14, lucas14, ewing15, malaska16}. Modeling \citep{tokano08} has indicated that the dunes require eastward flow to develop, however to attain atmosphere superrotation westward flow seems required. This poses a conundrum that might be resolved by a seasonal reversal of winds \citep{tokano08}. However, further investigation is warranted.

Again, Dragonfly will provide significant answers as to the dune properties, while improved remote sensing is required to search for time variations. Further laboratory work will also give comparative clues on the processes impacting the grains that comprise Titan's dunes.

\vspace{5mm}
\emph{12) When were the river channels carved and the low latitude seas filled?}

When the Huygens spacecraft descended to Titan's surface, it was finally revealed at the smallest scales \citep{tomasko05}, showing descent images of apparently empty river drainage networks leading to a dried-up seabed \citep{soderblom07, langhans12} (Fig~2d). A consensus is now emerging that Titan appears to be drying out over time, with formerly extensive low-latitude oceans of methane retreating polewards \citep{larsson13}. However it is important to know the timescale in which this has occurred, since it has important implications for our understanding of Titan's past history, and the unique nature of its atmosphere.

Given present-day rates of methane loss coupled with Titan's known topography, the equatorial seas are predicted to have been connected to polar seas 300 Myr ago, and a global ocean to have existed 600 Myr ago \citep{larsson13}. However, studies of river erosive timescales to form present topography are much less constraining, with estimates ranging from 60 Kyr to 470 Myr \citep{cartwright11}. In addition, material accumulated in alluvial fans may be dominated by infrequent, severe storms that quickly empty again, rather than more consistent river flow \citep{faulk17}, complicating the estimates of fluvial network creation timescales. 

To more fully understand the past history of liquids on Titan's surface, we must increase our knowledge in several areas. Firstly, an accurate and complete high-resolution topographic map is required, since topography is a major controlling factor of lake/sea location \citep{lora22} and fluvial network morphology \citep{miller21}. Temporal monitoring of shorelines is also necessary, to understand when and how changes are taking place at the present day, as seems to be the case at Ontario Lacus \citep{turtle11a}. Also, the atmosphere cannot be neglected here. In particular, having better measurements of the global methane humidity \citep{lellouch14, adamkovics07, adamkovics16} and its changes on seasonal timescales is important, to better constrain climate and meteorological models \citep[e.g.][]{friedson09, lebonnois12, mitchell16, lora19}.

\vspace{5mm}
\emph{13) Is there exposed water (or other) ice on the surface?}

Detection of exposed ices on Titan's surface would be an important finding (see also Chapter~11). Since much of the surface is covered by organics \citep[e.g.][]{lopes20} that apparently accumulate from the atmosphere \citep{clark10, rodriguez14, yu17a, yu18}, any exposed ices could be indicative of recent endogenic or exogenic activity, i.e. volcanism or impacts. In particular, any indication that Titan is active at the present day would have important implications for the sustenance of its atmosphere (Q7).

Early in the Cassini mission, excitement was generated by a reported detection of exposed carbon dioxide ice by the VIMS instrument \citep{barnes05}, although this was later disputed \citep{clark10}. However, the study of \citet{clark10} argued for a more robust spectral detection of benzene: somewhat surprisingly since benzene is a less abundant hydrocarbon in the atmosphere than others such as acetylene, which was not detected. Later, \citet{solomonidou18} proposed a latitudinal variation of Titan's surface composition, including exposed water ice as a major constituent at latitudes beyond $\pm$30{\dg} N and S. At this time, identification of specific surface ices has remained highly challenging and controversial.

The Dragonfly mission promises to yield considerable insights into the surface composition, through IR and mass spectroscopy, seismology, gamma ray-neutron probe, and other means. In particular, it will make a detailed survey of the large impact crater Selk \citep{barnes21, lorenz21}, where there may be regions of exposed ice \citep{solomonidou20}. However, completing a global picture of exposed ice on Titan's surface will require a dedicated orbiter \citep{mitri14a, rodriguez22} or perhaps some type of long-range aerorover, such as a balloon \citep{lorenz08c, coustenis09} or airplane \citep{barnes12}.

\subsection{Questions about Titan's atmosphere (14-20)}

\begin{figure}[htbp]
\centering
\includegraphics[width=1.0\textwidth]{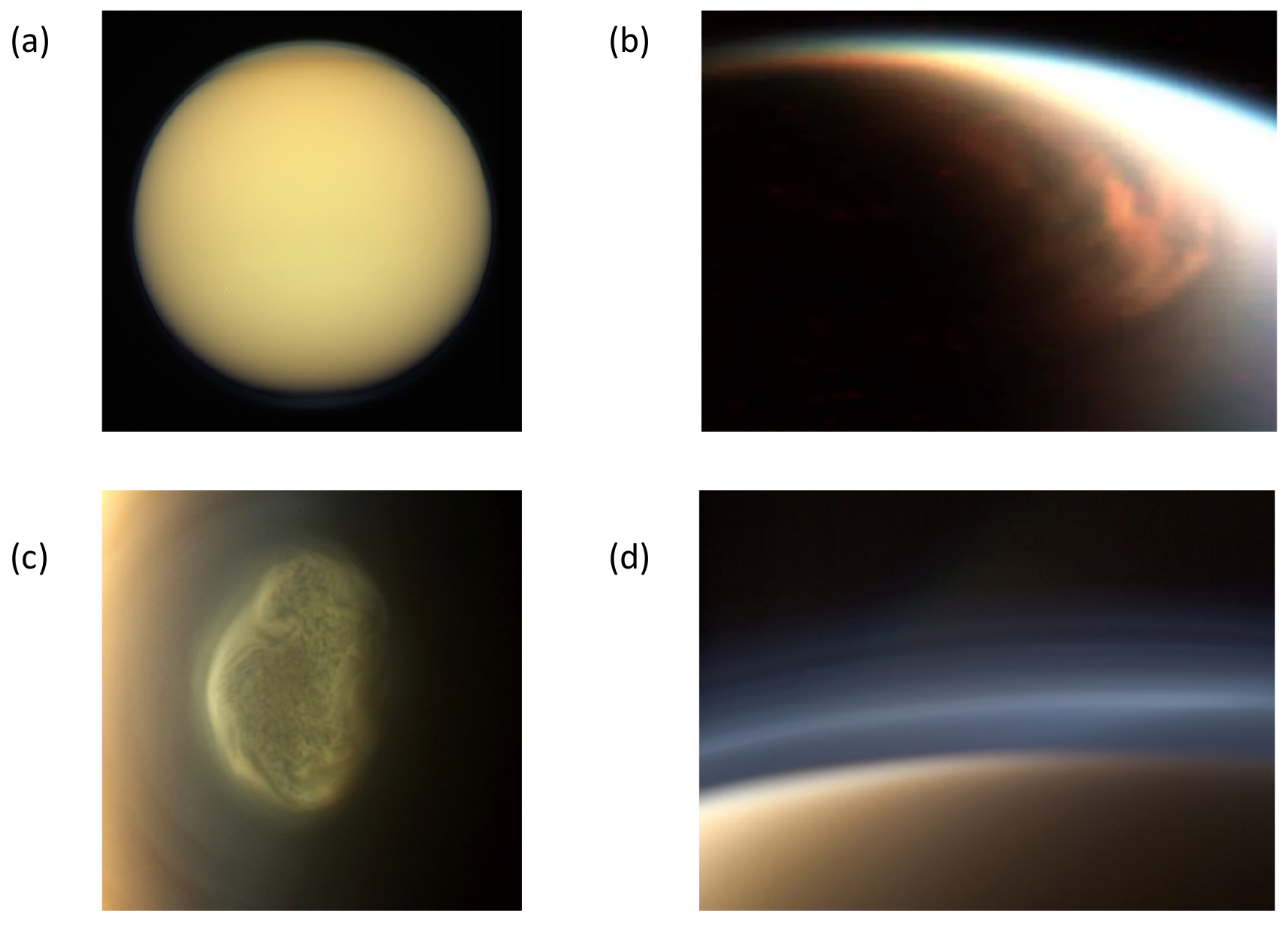} \\[0.2cm]
\caption{Views of Titan's atmosphere. (a) Distant Titan from Cassini ISS (12/16/2011), showing the ubiquitous golden haze (NASA/JPL-Caltech/SSI/Kevin M. Gill); (b) Ethane cloud over Titan's north pole from Cassini VIMS (NASA/JPL/University of Arizona/LPGNantes) - NASA image PIA09171; (c) Titan south polar ice cloud of HCN, June 27, 2012 (NASA/JPL-Caltech/SSI); (d) Titan limb haze showing detached haze layer Sept. 24, 2010 (NASA/JPL-Caltech/SSI/Jason Major). }
\label{fig:atmosphere}
\end{figure}

\vspace{5mm}
\emph{14) What types of clouds exist in the troposphere and stratosphere?}

Clouds on Titan (see Fig.~3b, 3c, Chapter 8) have been observed both directly, by imaging, and inferred indirectly by spectroscopy and modeling. The first evidence of large-scale clouds was obtained by brightening in the near infrared spectrum noticed by \citet{griffith98, griffith00}, and subsequently imaged using ground-based telescopes with adaptive optics \citep{brown02,brown09,brown10, schaller06a, schaller06b, bouchez04, roe04}. These were inferred to be clouds of methane in the upper troposphere.

Cassini imaging and spectroscopy revealed yet more clouds, including not only further instances of tropospheric clouds \citep{rodriguez09, turtle09, turtle11b, turtle11c, turtle18}, but also stratospheric ice clouds: a polar ethane cloud in late northern winter \cite{griffith06b}, north polar cyanoacetylene and dicyanoacetylene ice clouds \citep{anderson10, anderson16}, a south polar hydrogen cyanide cloud \citep{dekok14}, and a benzene ice cloud \citep{vinatier18}. The existence of many condensible species in Titan's atmosphere implies that cloud composition can be more complex still, likely with layering of ices accumulated at different altitudes \citep{anderson16}.

A new front in cloud research has appeared in recent years due to the laboratory investigation of organic co-crystals, co-condensed phases of one or more molecules \citep{cable14, cable18, cable19, cable20, cable21a, czaplinski23, maynard-casely16, vu14, vu20a, thakur23, francis23, ennis20}. Understanding of these exotic substances is helping to inform our understanding of stratospheric cloud formation and growth and the materials that may be on Titan's surface.

Current knowledge of Titan's cloud composition and occurrence is spotty, due to a limited amount of high-resolution imagery and spectroscopy during Titan close flybys, combined with sporadic monitoring from the Earth. Given the complexity of cloud scenarios on Titan, a much more sustained and regular monitoring is necessary, together with cloud particle growth experiments in Earth laboratories.

\vspace{5mm}
\emph{15) How often does it rain on Titan? }

Although we can be certain about clouds in Titan's lower atmosphere, rainfall is more difficult to observe directly, and models include some amount of rainfall that evaporates before reaching the ground (known as `virgae') \citep{barth07,graves08}.  However, indirect evidence for rain exists through surface darkening observed after the passage of a large storm front, followed by a lightening of the surface observed some weeks later, apparently due to re-evaporation \citep{turtle11b}.

Better knowledge of the actual incidence and intensity of precipitation events is valuable however, since it has important consequences for pluvial and fluvial erosion \citep{lorenz96, jaumann08, lorenz08d, legall10, cartwright11, langhans12, black12, tewelde13, burr13a, aharonson14, moore14, howard16, neish16, faulk17, radebaugh18, matteoni20, miller21, lewis-merrill22}, and the rate of replenishment of liquid bodies (lakes and seas) \citep{sotin12}. In addition, knowing the latitudinal and temporal pattern of rainfall can help to explain the observed dry equatorial regions and the hemispheric asymmetry in lakes and seas \citep{aharonson09, lora15, tokano19, lora22, mitchell08, mitchell09, mitchell16, schneider12, birch18}.

Collecting such information will not be easy, requiring a sustained observational campaign best conducted from Titan orbit, as well as potentially observations from the surface, and ideally in-situ sampling, for example by high-altitude balloons \citep{lorenz08c, coustenis09}. Further modeling work will also help to determine whether observed rainfall can be adequately explained.

\vspace{5mm}
\emph{16) What is the availability of P and S compounds for chemistry?}

Six elements are considered the most vital for life on Earth: H, C, N, O, P and S ('CHNOPS'), together constituting 98\% of living matter \citep{cockell16}. To date four of the six have been found on Titan (H, C, N, O), for example in hydrogen cyanide and carbon monoxide, while all six have been found or strongly suspected in Enceladus plume material \citep{waite09, cable21b, postberg23}. Therefore, determining whether there is available phosphorus and sulfur on Titan assumes an astrobiological significance \citep{raulin12}, as well as a geochemical one \citep{fortes07, pasek11}. Searches have been made for atmospheric hydrogen sulfide, carbon sulfide and phosphine with null result so far \citep{nixon13b, teanby18}. 

Attempts to detect P and S in the atmosphere can continue to be made by astronomical observations and remote sensing, e.g. with ALMA, IRTF etc; while the Dragonfly mission in the 2030s will be able to sample both the atmosphere and surface with excellent ability through its mass spectrometer to detect any P and S compounds. 

\vspace{5mm}
\emph{17) What is the composition of Titan's aerosols?}

Understanding Titan's organic particulate aerosols - which give rise to its famous golden appearance (Fig.~3a) and multiple haze layers (Fig.~3d)  - is important for many reasons (see also Chapters 6 and 7). First, these are critical in the radiative balance of Titan's atmosphere, including regulation of its `greenhouse effect' \citep{mckay91}. Second, the chemical complexity of the aerosols is vital for understanding the astrobiological potential of Titan's atmosphere \citep{khare86, neish10, horst12, kawai13, cleaves14, he14, kawai19, gautier14}. Third, they serve as condensation nucleii for cloud particles and droplets \citep{mckay01, horst13, burgalat14}. And fourth, the aerosols undoubtedly sediment onto the surface, where they may form a significant component of the equatorial dunes \citep{barnes08, barnes15, yu17a}.

Since the 1980s there has been a sustained effort to replicate Titan's atmospheric haze particles through laboratory experiments, in which vacuum vessels filled with varying mixtures of nitrogen , methane and other gases are reacted together by energetic stimulation with UV or electric current \citep[see review by][]{cable12}. Their fractal nature has been elucidated by modeling and SEM techniques \citep{cabane93, rannou95, rannou03, rannou99, bar-nun08, schulz21}.

During their atmospheric descent, aerosols are moreover predicted to chemically and physically evolve \citep{lavvas11a, carrasco18, perrin21}. In the ionosphere, aerosols have been shown to further influence their atmospheric environment \citep{chatain23a}, with an impact on the electron populations detected by the Cassini RPWS/LP instrument \citep{chatain21}.   

A direct attempt was made to sample aerosol material with Huygens' ACP instrument \citep[Aerosol Collector Pyrolyzer,][]{israel03}, which was designed to pass pyrolysis products to the mass spectrometer \citep[GCMS - Gas Chromatograph and Mass Spectrometer,][]{niemann02}. The results showed significant amounts of hydrogen cyanide and ammonia \citep{israel05, coll13}.

Sampling and elucidating the aerosol composition and physical properties is a major goal for future missions \citep{mitri21, rodriguez22}, since the aerosols are one of Titan's most distinguishing features. The DraMS instrument on Dragonfly will be able to measure the composition of surface organics from the dune fields \citep{trainer21}, however direct sampling of atmospheric aerosols remains a future goal. In addition, instruments other than MS such as microscopy can play a role in illuminating the physical properties. Further laboratory work can expand our concept of the aerosol parameter space, and also the reactions that they can undergo as nanoparticles in the ionosphere until they sediment as submillimetric particles at Titan's surface.

\vspace{5mm}
\emph{18) What is the cause of the tilted atmospheric rotation?}

Early results from mapping of Titan's atmospheric temperature contours by Cassini CIRS revealed an unexpected phenomenon: the axis of the atmospheric rotation pole was offset by some $\sim$3.5\dg\ from the solid body rotation axis \citep{achterberg08b}. Later studies confirmed this finding \citep{roman09, teanby10c, cordiner19a, sharkey20, kutsop22}. To date, an explanation for this phenomenon has remained elusive \citep{nixon18}, and further climate simulations are required to better understand how it occurs and whether it is changing over time.

\vspace{5mm}
\emph{19) Are there important biological precursor molecules produced in the atmosphere?}

The production of hydrocarbons and nitriles in Titan's atmosphere was predicted by photochemical models as far back as the 1970s \citep{strobel74}, validated soon thereafter by mid-infrared spectroscopy from Voyager 1, which revealed a rich tapestry of molecular vibrations throughout the mid-infrared \citep{hanel81, maguire81, kunde81}. Since then, the known molecular inventory has been expanded repeatedly, e.g. by ISO \citep{coustenis98, coustenis03}, and most revealingly by direct sampling with Cassini INMS and CAPS \citep{niemann05, waite05, waite07, vuitton07, niemann10} - see also Chapter 7.

Laboratory investigations have shown that simulated Titan photochemical products, produced by reactions starting with nitrogen and methane reagents, can be hydrolyzed to produce amino acids in a manner similar to a Miller-Urey experiment \citep{khare86, neish10, cleaves14}. Other experiments showed that such amino acids can even be produced directly in the atmosphere when oxygen is included in the reactions by admission of carbon monoxide \citep{horst12}.

The combination of the detected complex organic chemistry along with the laboratory simulations showing the potential for prebiotic molecules to arise has naturally led to the question of whether such prebiotic molecules also occur on Titan. Despite some of Cassini's instruments (especially CAPS) being able to detect the presence of large organic molecules, the instruments did not have the ability determine the molecular composition or structure.

In future, the search for evidence of biological precursor molecules on Titan can proceed on two fronts. First, astronomical measurements will continue to push the boundaries of the known atmospheric inventory through infrared and sub-millimeter spectroscopy \citep[e.g.][]{cordiner15, palmer17, lombardo19c, thelen20, nixon20}. This will be enabled by upgraded capabilities at existing observatories and entirely new telescopes on the ground and in space (see Section 3 of this chapter),  backed up by additional laboratory investigations of gases and ices at Titan relevant temperatures. 
In addition, in situ measurements are needed with a more sophisticated chemical analyzer: this will be achieved by the DraMS instrument onboard Dragonfly \citep{barnes21, trainer21}.

\vspace{5mm}
\emph{20) Is Titan's atmosphere stable over time?}

Question 3 above described the uncertainties regarding the origin of Titan's atmosphere: a closely linked question is about its fate. Much depends on whether methane is being resupplied to the atmosphere or not (Chapter 13), as well as its rate of photochemical depletion (see Chapters 6 and 7). Proposed mechanisms of endogenous resupply include: destabilization of crustal clathrates \citep{tobie06}; displacement of surficial clathrates \citep{choukroun12}; and geochemical production in the interior from \coo\ through serpentinization \citep{glein15}, although \citet{fortes12} suggests that deep interior methane will be captured in clathrate and not escape to the atmosphere.

On the other hand, if methane is being irreversibly depleted, then eventually its current methane greenhouse will disappear, being replaced by a much cooler climate under a nitrogen greenhouse \citep{charnay14}, or the atmosphere may collapse altogether into a `snowball' state \citep{lorenz97a, wong15}.

Would such a condition persist forever? Impacts may work to reverse the snowball state by vaporizing enough methane into the atmosphere to reconstitute something resembling the warmer state we see today; alternatively Titan under a future red giant Sun will have a dramatically different climate \cite{lorenz97b}.

Predicting the future state of the atmosphere by climate modeling requires most critically a firmer understanding of whether methane is being resupplied to the atmosphere by some mechanism \citep[e.g.][]{tobie06, glein15} or not. Closing this loophole in our knowledge in turn will hinge on a combination of investigative efforts: determining if the atmosphere has already changed over time, perhaps by sensitive isotopic measurements and modeling, and searching for methane resupply by cryovolcanism \citep[surface changes,][]{solomonidou16}, or buffering by the solid \citep[clathrates,][]{davies16} or liquid surface \citep[lakes and seas,][]{hayes16}. 

Instrumentation that could help guide us to these goals includes: ice penetrating RADAR; visible and near-IR imagery and spectroscopy (spectro-imagery), mass spectroscopy for improved measurements of isotopic ratios and noble gas abundances \citep{coustenis09, tobie14, mitri21, sulaiman21, rodriguez22}. Therefore, further orbital and {\em in situ} investigation is necessary.

\section{Future Directions}

\begin{figure}[htbp]
\centering
\includegraphics[width=1.0\textwidth]{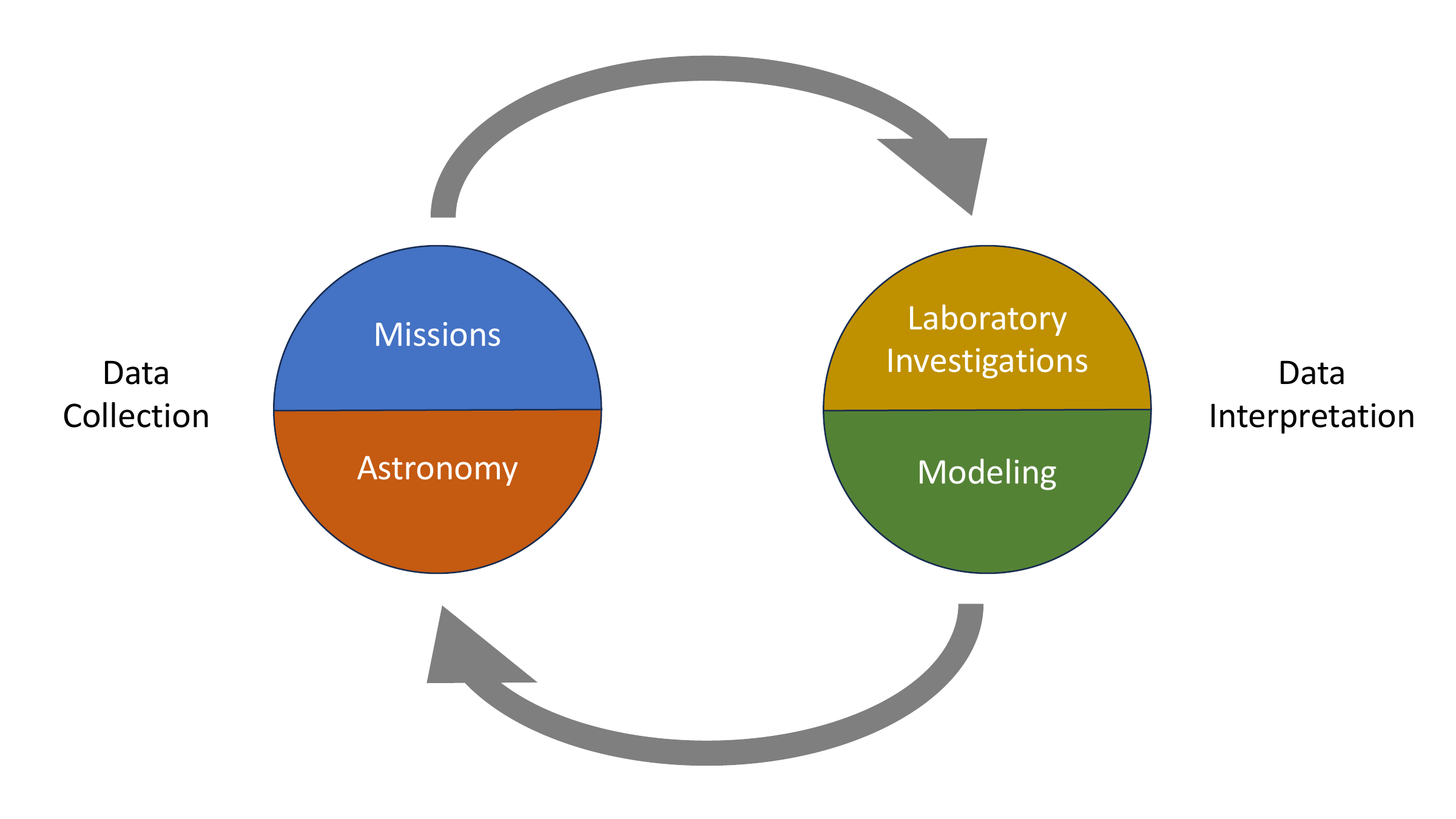} \\[0.2cm]
\caption{Synergies between data collection (astronomy, missions) and interpretation (modeling, laboratory investigations). None of these techniques works in isolation, but serve to reinforce each other to derive new knowledge.}
\label{fig:synergies}
\end{figure}

We conclude this chapter - and book - by mentioning again some of the means by which the questions raised in this chapter may be answered in future. The categorial divisions here echo the four main techniques introduced at the start of the chapter (astronomy, laboratory work, modeling, missions), which work synergistically together to collect and interpret data (Fig.~\ref{fig:synergies}). In addition, we propose the increased use of a fifth type of investigation: terrestrial field analog studies.

\begin{figure}[htbp]
\centering
\includegraphics[width=1.0\textwidth]{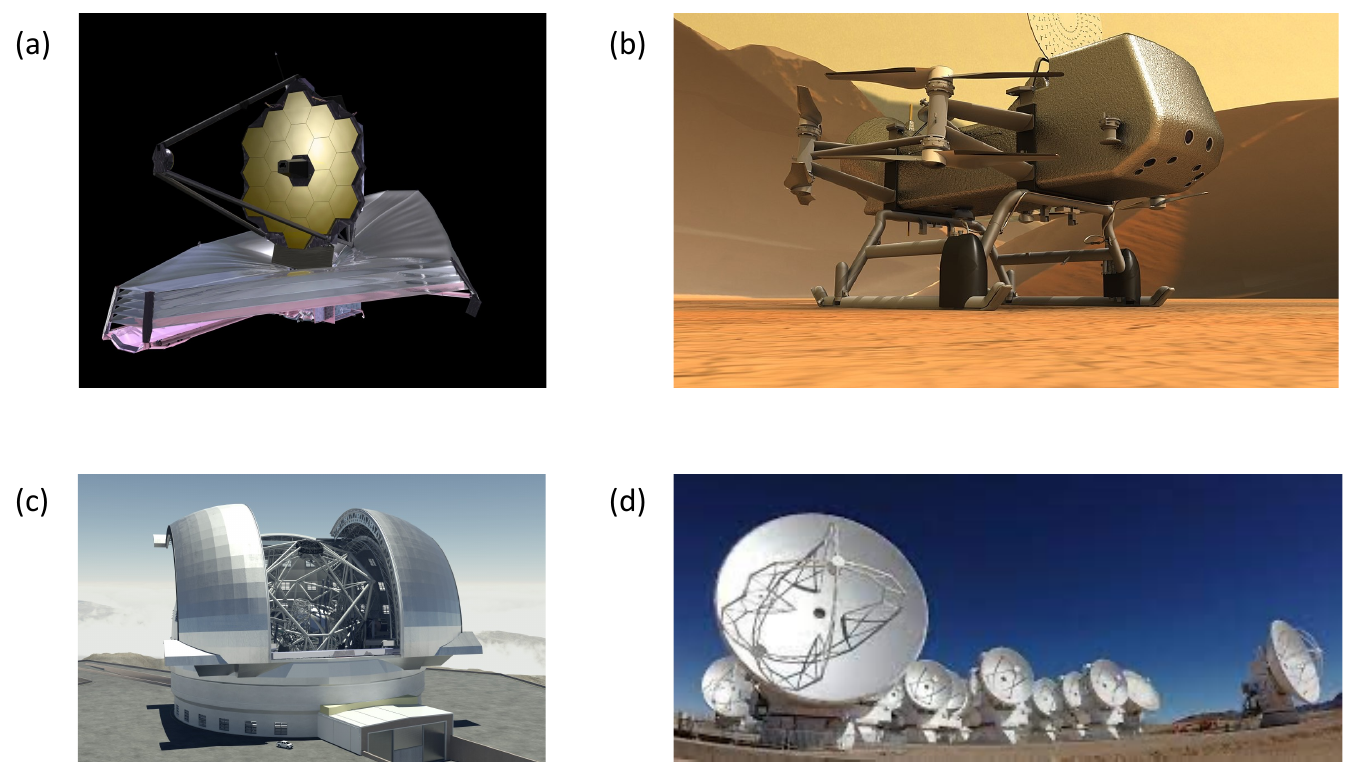} \\[0.2cm]
\caption{Future means of probing Titan. (a) NASA's James Webb Space Telescope launched 12/15/2021 (image: NASA); (b) NASA's Dragonfly mission to Titan scheduled for launch in 2028 (NASA/APL); (c) Artists impression of the completed 39m Extremely Large Telescope (ELT) being built by the European Southern Observatory (ESO), scheduled for first light $\sim$2029 (Swinburne Astronomy Productions/ESO); (d) Atacama Large Millimeter/submillimeter Array (ALMA). }
\label{fig:future}
\end{figure}

\vspace{5mm}
\emph{Telescopes}

As mentioned at the opening of this chapter, ground and (later) space-based telescopes have played an important role in unraveling Titan's mysteries.\footnote{Indeed, without the invention of the telescope we not know of its existence at all. } This has included spectroscopy to determine its gases, imaging with adaptive optics to view the surface and clouds, and synthetic aperture interferometry for imaging and spectroscopy at longer wavelengths. 

At the time of writing, the James Webb Space Telescope (JWST, Fig.~\ref{fig:future}(a)) is the newest of the large, general purpose space telescopes to come on-line, and has already begun to make execute a planned range of spectroscopic and imaging observations of Titan \citep{nixon16a}. On the ground, many optical/IR telescopes are routinely used to make observations of Titan, such as cloud tracking \citep{griffith00, brown02, roe02, depater06, schaller06a, schaller06b, adamkovics07, schaller09, brown09, adamkovics16, adamkovics17}, spectroscopy \citep{roe03, lellouch03, adamkovics04, penteado05, penteado10, adamkovics16, lombardo19b}
and occultation observations \citep{hubbard90, sicardy90, sicardy99, sicardy06, zalucha07}, including large telescopes with adaptive optics such as Keck, the Very Large Telescope (VLT), Gemini and others. In future, even larger telescopes currently planned such at the Giant Magellan Telescope \citep[GMT, 25~m,][]{fanson18, fanson20, fanson22} the Thirty Meter Telescope \citep[TMT,][]{nelson08, sanders13, skidmore15} and the Extremely Large Telescope \citep[ELT, 39~m,][]{gilmozzi07,tamai14} (see Fig.~\ref{fig:future}(c)) will greatly improve our ability to investigate Titan's atmosphere and surface from Earth.

Large surface interferometric arrays such as the Karl Jansky Very Large Array (VLA), the Very Long Baseline Array (VLBA), the Atacama Large Millimeter-submillimeter Array (ALMA, Fig.~\ref{fig:future}(d)), and the Northern Extended Millimeter Array (NOEMA) are making important observations at long wavelengths \citep{muhleman90, butler04, witasse06, cordiner14, serigano16, molter16, lai17, palmer17, thelen18, lellouch19, nixon20}, and are poised to be upgraded in coming decades with wider receiver bands, improved sensitivity and longer baselines. 

\vspace{5mm}
\emph{Laboratory Work}

Data derived from observations and measurements of Titan must be interpreted, and observations and instruments must be carefully designed to maximise science. Laboratory work in many varieties has proven essential in this regard, as mentioned earlier in this chapter. In future, many more laboratory investigations are necessary including: high-resolution gas phase spectroscopy of molecules; spectroscopy of ices, especially mixed phases; creation and analysis of analogs (tholins) under a wide variety of conditions; erosional experiments at realistic Titan conditions; gas reaction and gas-particle reaction kinetics; and many others besides. An important area of current work is achieving consistency between results obtained at different laboratories, which has begun to motivate intercomparison efforts \citep{li22b}.

\vspace{5mm}
\emph{Modeling}

Modeling has always played a crucial role in our understanding of Titan, including photochemical models of the atmosphere; interior structure models; formation and dynamical evolution models; cloud models and GCMs; and others. With ever-increasing computing power, our capability to run models at higher levels of fidelity is always increasing (e.g. reducing grid size in GCMs, running photochemical models with larger networks and over more timesteps). In tandem, better constraints are needed for these models, including reaction rates, measured temperature and wind fields, seasonal observations of gas distributions, and more. Given the very large of possible reactions that can be included in photochemical models, work to define the most important pathways by Monte Carlo approaches is especially critical \citep{hebrard05, hebrard06, hebrard07}, to allow focusing of laboratory and theoretical work on the most influential reactions.

\vspace{5mm}
\emph{Missions}

Despite all the advances that can be made on Earth and using remote observations, the ultimate treasure trove of Titan knowledge is Titan itself. Hence, there is often no substitute for in situ measurements, covering everything from atmospheric temperature, pressure and gas composition, to surface morphology, land and sea composition, and much more. To this end, Titan scientists are excited about the upcoming Dragonfly mission (Fig.~\ref{fig:future}(b)) which will make a host of in situ measurements of Titan's dune fields and the environs of Selk crater. Even more ambitious missions have been proposed including orbiters, balloons, airplanes, boats and submarines \citep{coustenis09, stofan13, barnes12, mitri14a, tobie14, oleson15, nixon16b, nixon19, rodriguez22, barnes22}. 

\vspace{5mm}
\emph{Terrestrial Field Analog Studies}

Historically, terrestrial field studies of planetary analog environments have proven highly beneficial for both interpreting returned data (especially lunar and martian imagery) but also for testing instrumentation and the training of astronauts. To date terrestrial field studies targeting Titan analog sites have been limited and primarily focused on dune analogs \citep{radebaugh10}. However, the types of useful terrestrial analog environments for Titan may in fact be much larger (e.g. ices, rivers, lakes, seas, karst, polar stratospheric clouds, to name a few), providing fertile new avenues for research.

\vspace{5mm}
\noindent
{\bf Acknowledgements}

The authors wish to thank the editors of the book, Dr Rosaly Lopes and Dr Anezina Solomonidou, along with three referees - Dr Ralph Lorenz and two anonymous reviewers - for helpful feedback provided during the preparation of this chapter. 

\bibliography{allpapers}
\bibliographystyle{elsarticle-harv}

\end{document}